\documentclass[twocolumn,showpacs,preprintnumbers,amsmath,amssymb]{revtex4}
\font\mybb=msbm10 at 10pt
\def\bb#1{\hbox{\mybb#1}}

\def\bR {\bb{R}}

\newcommand{\be}{\begin{equation}}
\newcommand{\ee}{\end{equation}}
\newcommand{\bea}{\begin{eqnarray}}
\newcommand{\eea}{\end{eqnarray}}
\newcommand{\ba}{\begin{array}}
\newcommand{\ea}{\end{array}}

\def\bbox{{\,\lower0.9pt\vbox{\hrule \hbox{\vrule height 0.2 cm
\hskip 0.2 cm \vrule height 0.2 cm}\hrule}\,}}
\newcommand{\dsl}{\pa \kern-0.5em /}

\font\mybb=msbm10 at 10pt
\def\bb#1{\hbox{\mybb#1}}

\def\bR {\bb{R}}

\def\bX {\bb{X}}
\def\bP {\bb{P}}

\def\bU {\bb{U}}
\def\bV {\bb{V}}

\usepackage{graphicx}% Include figure files
\usepackage{dcolumn}% Align table columns on decimal point
\usepackage{bm}% bold math

\begin{document}

\preprint{UMTG-17 , DAMTP-2010-55}

% Force line breaks with \\

\title{\large Anyons from Strings 
}

\author{Luca Mezincescu$^\star$  and Paul K. Townsend$^\dagger$}

\affiliation{$\star$  Department of Physics, University of Miami,
Coral Gables, FL 33124, USA \\
$\dagger$ Department of Applied
Mathematics and Theoretical Physics, \\
Centre for Mathematical Sciences, Univ.  of Cambridge, \\
 Wilberforce Road, Cambridge CB3 0WA, UK}


\begin{abstract}

The Nambu-Goto string  in a 3-dimensional (3D) Minkowski spacetime  is quantized preserving Lorentz invariance and parity.  The spectrum of massive states contains anyons.  An ambiguity in the ground state energy  is resolved by the 3D ${\cal N}=1$ Green-Schwarz superstring, which has massless ground states describing a dilaton and dilatino, and first-excited states of spin 1/4.

%An article usually includes an abstract, a concise summary of the work
%covered at length in the main body of the article. It is used for
%secondary publications and for information retrieval purposes. Valid
%PACS numbers may be entered using the \verb+\pacs{#1}+ command.
\end{abstract}

\maketitle

A standard claim of  string theory texts is that a free string cannot be consistently quantized below its critical dimension, preserving Lorentz invariance,  without the introduction of  an additional  ``Liouville'' variable that is absent classically.  Here we show that  strings in a three-dimensional (3D) spacetime are an exception to this rule.  Specifically, we show that both the 3D Nambu-Goto string and  the
${\cal N}=1$ 3D Green-Schwarz (GS) superstring \cite{Green:1983wt} may be quantized, preserving both Lorentz invariance and parity, without the introduction of any additional variables. It turns out that the spectrum of these strings includes anyons, i.e. particles of spin $s$ such that $2s$ is not an integer \cite{Binegar:1981gv,Wilczek:1981du}.  Specifically, we find anyons  in the bosonic string spectrum at level $2$ or $3$, depending on the choice of ground state energy. The superstring is massless in its ground state, has spin $1/4$ at level $1$ and other ``semions''  (particles of spin $1/4 + n/2$ for integer $n$) at  level $2$. 

One may ask why the existence of these new 3D quantum strings  has not previously been noticed (we do not say ``string theories'' because we do not address here  issues of modular invariance or interactions).  Part of the answer to this question is surely that a manifestly covariant description of   anyons requires fields in representations of some multiple cover of $Sl(2;\bR)$. The universal cover is required for irrational spin (and potentially for an infinite number of rational spins)  implying infinite-component fields  \cite{Jackiw:1990ka,Plyushchay:1990rt}. However, such  fields do not arise in any of the standard approaches to covariant  quantization. Although this limitation may be circumvented in the future, at present it is only in the light-cone gauge that one can easily see all possibilities for a consistent quantum  theory, and that is the method used here. 

We begin with the Hamiltonian form of the Nambu-Goto action for a closed relativistic 3D string of tension $T$ in terms of the canonical 3-vector variables $(\bX, \bP)$, which  are functions of the worldsheet time 
$\tau$ and the string coordinate $\sigma \sim \sigma + 2\pi$:
 \be\label{closedstring}
S =  \!\!\int\! \! d\tau\! \oint\! \!\frac{d\sigma}{2\pi} \!\left\{\dot {\bX}^\mu {\bP}_\mu
- \frac{1}{2}\ell \left[{\bP}^2 + (T{\bX}^\prime)^2\right]     
 - u\,  {\bX}^{\prime\mu} {\bP}_\mu \right\}. 
\ee
The overdot and prime indicate derivatives with respect to $\tau$ and $\sigma$,  respectively,  and  $\ell$ and $u$ are Lagrange multipliers  for the Hamiltonian and string-reparametrization constraints, respectively.  This action involves the Minkowski spacetime metric  (with `mostly plus' signature)  via the scalars $\bP^2$ and $({\bX}^\prime)^2$.  The standard Nambu-Goto action is recovered by elimination of the 3-momentum $\bP$ followed by elimination of $\ell$ and then $u$. In addition to the gauge invariances associated to the constraints, the action is invariant under the Poincar\'e transformations generated by the Noether charges
\be\label{noether}
{\cal P}_\mu = \frac{1}{2\pi}\oint\! d\sigma\,  {\bP}_\mu \, , \qquad 
{\cal J}^\mu = \frac{1}{2\pi}\oint\! d\sigma \left[ {\bX}\wedge {\bP}\right]^\mu\, , 
\ee
where $\left[{\bU}\wedge{\bV} \right]^\mu = \varepsilon^{\mu\nu\rho} {\bU}_\nu{\bV}_\rho$ for any two 3-vectors $\bU$ and $\bV$, and the invariant antisymmetric tensor $\varepsilon$  tensor is defined such that  $\varepsilon^{012}=1$. 

We now introduce light-cone coordinates
\be
X^\pm = \frac{1}{\sqrt{2}}\left({\bX}^1 \pm {\bX}^0\right) \, ,\qquad 
P_\pm = \frac{1}{\sqrt{2}}\left({\bP}_1 \pm {\bP}_0\right)\, , 
\ee
and set $\bX^2 = X$ and $\bP_2=P$.  It is convenient to define
\begin{eqnarray}
x(\tau) &=&\frac{1}{2\pi} \oint \! d\sigma X\, , \qquad  x^-(\tau) = \frac{1}{2\pi}\oint \! d\sigma X^-\, ,\nonumber \\
p(\tau) &=&  \frac{1}{2\pi}\oint \! d\sigma P\, , \qquad p_+(\tau) =  \frac{1}{2\pi}\oint \! d\sigma P_+
\end{eqnarray}
and 
\begin{eqnarray}
\bar X &=& X-x\, , \qquad \bar X^- = X-x^- \, , \nonumber\\
\bar P &=& P-p \, ,  \qquad \bar P_+  = P_+ - p_+ \, .
\end{eqnarray}
The light-cone gauge is defined by the choice
\be\label{lcg}
X^+ =\tau \, , \qquad P_- = p_-(\tau)\, , 
\ee
where $p_-(\tau)$ is a non-zero function of $\tau$ only.  This choice leaves only the  residual global gauge invariance  that shifts the origin of the angular string coordinate $\sigma$.   In this gauge, the Hamiltonian constraint imposed by $\ell$ may be solved for $P_+$:
\be\label{lapsecon}
P_+ = -\frac{1}{2p_-} \left[ P^2 + (TX^\prime)^2\right]\, . 
\ee 
One also finds that $\bar X^-$ is a Lagrange multiplier imposing the constraint $u^\prime=0$, which has the solution $u=u_0(\tau)$. The final result is the Lagrangian
\be\label{lcgLag}
L = \left\{\dot x p  + \dot x^- p_- + \oint \! \frac{d\sigma}{2\pi} \, \dot {\bar X} \bar P \right\} -H 
- u_0 \oint \! \frac{d\sigma}{2\pi}\,  \bar X' \bar P \,  , 
\ee
where the Hamiltonian is 
\be\label{Ham}
H = -p_+ = \frac{1}{2p_-} \left[ p^2 + 
\oint \! \frac{d\sigma}{2\pi}  \left\{ \bar P^2 + (T{\bar X}^\prime)^2\right\} \right]\, . 
\ee
As expected, there is a residual global constraint imposed by $u_0$.  
In the light-cone gauge, the Poincar\'e Noether charges of (\ref{noether}) are
\begin{eqnarray}\label{LorentzString}
{\cal P} &=& p\,  , \qquad {\cal P}_- = p_- \qquad {\cal P}_+ = -H\, , \nonumber \\
{\cal J} &=&  x^- p_- + \tau H\, , \qquad {\cal J}^+ = \tau p  -x p_-\, \\
{\cal J}^- &=&  -x^- p -xH  + \oint\! \frac{d\sigma}{2\pi} 
\left[ \bar X {\bar P}_+ - \bar X^- \bar P\right] \, .  \nonumber 
\end{eqnarray}
The two Poincar\'e invariants are
\begin{eqnarray}\label{poincinv}
-{\cal P}^2 &\equiv& {\cal M}^2 = 
\oint \! \frac{d\sigma}{2\pi}  \left[ \bar P^2 + (T{\bar X}^\prime)^2\right]\, , \nonumber \\
{\cal P} \cdot{\cal J} &\equiv& \Lambda = p_- \oint\! \frac{d\sigma}{2\pi} 
\left[ \bar X {\bar P}_+ - \bar X^- \bar P\right] \, . 
\end{eqnarray}

We now Fourier expand the canonical pair $(\bar X, \bar P)$ by writing
\begin{eqnarray}\label{separate}
\bar P -T\bar X^\prime &=& \sqrt{2T}\sum_{n=1}^\infty \left[e^{in\sigma} \alpha_n + e^{-in\sigma}  \alpha_n^*\right] \, , \nonumber \\
\bar P +T\bar X^\prime &=& \sqrt{2T}\sum_{n=1}^\infty \left[e^{in\sigma} {\tilde \alpha}_n^* + e^{-in\sigma} {\tilde \alpha}_n\right]\, . 
\end{eqnarray}
The Lagrangian (\ref{lcgLag}) becomes
\begin{eqnarray}\label{oscLag}
L &=& \dot x p  + \dot x^- p_- + i \sum_{n=1}^\infty n^{-1}\left( \dot \alpha_n \alpha_n^* + \dot{\tilde \alpha}_n \tilde \alpha_n^* \right) \nonumber \\
&&  \!\!\!\!\!\! \!\! -\frac{1}{2p_-} \left( p^2 + {\cal M}^2\right) + 
u_0 \sum_{n=1}^\infty \left(\alpha_n^* \alpha_n - \tilde \alpha_n^*\tilde \alpha_n\right)\, , 
\end{eqnarray}
where
\be
{\cal M}^2 =  2T \sum_{n=1}^\infty \left( \alpha_n^* \alpha_n + \tilde \alpha_n^*\tilde \alpha_n\right)\, .
\ee

Observe that the other Poincar\'e invariant $\Lambda$ of  (\ref{poincinv})  depends on $\bar X^-$ as well as the canonical variables of the final action, but the  equation of motion of  $u$ in the original action reduces in the light-cone gauge to
\be\label{Xminuseq}
p_- (\bar X^-)^\prime = - \bar X^\prime P\, ,
\ee
which  allows us to express $\bar X^-$ in terms the Fourier coefficients of 
$(\bar X,\bar P)$.  
We pass over the details, which are similar to those for the critical string \cite{Goddard:1973qh}; the final result is
\be\label{Lambdalambda}
\Lambda =  \sqrt{2T}\left( \lambda + \tilde\lambda\right)\, ,
\ee
where $\lambda$ depends only on the $\alpha_n$  and $\tilde\lambda$ is the same expression but in terms of the $\tilde\alpha_n$.  Explicitly, 
\be
\lambda=  \sum_{n=1}^\infty \frac{i}{n} \left(\alpha_n^* \beta_n - \alpha_n\beta_n^* \right) \, , 
\ee
where
\be
\beta_n = \frac{1}{2}\sum_{m=1}^{n-1} \alpha_m \alpha_{n-m} + \sum_{n>m} \alpha_m\alpha_{n-m}^*\, ,
\ee
and similarly for $\tilde\lambda$.

To quantize we promote the canonical variables to operators satisfying canonical commutation relations. The non-zero commutators are
\be
[x^-,p_-] = [x,p] =i\, \quad [\alpha_n ,\alpha_n^\dagger] =
 [\tilde\alpha_n ,\tilde\alpha_n^\dagger] =n\, . 
\ee
The constraint imposed by $u_0$ becomes the level-matching condition in the quantum theory:
\be
N= \tilde N\, , \qquad N= \sum_{n=1}^\infty \alpha_n^\dagger \alpha_n,  \quad{\tilde N}= \sum_{n=1}^\infty {\tilde\alpha}_n^\dagger {\tilde \alpha}_n\, .
\ee
Taking this into account, the mass-squared operator is
\be
{\cal M}^2 = 2T \left(2N-a\right)\, , 
\ee
where $a$ is an arbitrary constant arising from operator ordering ambiguities. 
Similarly, the operator $\Lambda$ is as in (\ref{Lambdalambda}) but now with
$\alpha_n^*\to \alpha_n^\dagger$ and hence $\beta_n^*\to \beta_n^\dagger$, and 
similarly for $\tilde \lambda$. 

The quantum Lorentz generators are 
\begin{eqnarray}
{\cal J} &=& \frac{1}{2}\left\{x^-, p_-\right\} + \tau H \, , \qquad {\cal J}^+ = \tau p -x p_-\, , \nonumber \\
{\cal J}^- &=&  -x^- p - \frac{1}{2}\left\{x,H\right\} + \Lambda/p_-\, .
\end{eqnarray}
In principle, there are operator ordering ambiguities in these expression  but they are fixed by the requirements of hermiticity and closure of the Lorentz  algebra. It is straightforward to verify that  the charges as given satisfy the required commutation relations:
\be
\left[{\cal J}, {\cal J}^\pm\right]  = \pm i {\cal J}^\pm \, , \qquad \left[{\cal J}^+,{\cal J}^-\right] = i {\cal J}\, . 
\ee
This should not be a surprise because the ``dangerous'' commutators vanish `by default'  in three dimensions. 

Because the Poincar\'e invariant operators ${\cal M}^2$ and $\Lambda$ commute, they may be simultaneously diagonalized.  It follows that the space spanned by the level-$N$ states is an invariant subspace of $\Lambda$. At levels $N=0,1$ there is a single state that is annihilated by $\Lambda$, 
so non-zero eigenvalues of $\Lambda$ can occur only for  $N\ge2$. The eigenvalues of $\Lambda$ at each level  divided by the mass of the level are the ``relativistic helicities'' at that level. The four states at level $2$ have helicities 
\be
\left(0,0, \pm \tfrac{3}{\sqrt{4-a}}\right)\, ,  
\ee
which implies that $a<4$. The $9$ states at level 3 have helicities 
\be
\left(0,0,0, \pm \sqrt{\tfrac{179}{12(6-a)}}, \pm \sqrt{\tfrac{179}{12(6-a)}}, 
\pm \sqrt{\tfrac{179}{3(6-a)}}\right)\, .
\ee
Observe that non-zero helicities appear in parity doublets of opposite helicity, so parity is preserved by the quantization.   For any $a\le4$  there is an anyon in either level $2$ or level $3$.  It would be natural to choose $a=0$, so that the level-$0$ state is massless; in this case we have a massive scalar at level $1$, spin $3/2$ at level $2$ and irrational spin anyons at level $3$. Another  natural choice is $a=2$ because this makes the level-$0$ state a  tachyon and the level-$1$ state a massless scalar, which might be interpretable as a dilaton, as for the critical bosonic string; in this case there is a massive state with irrational spin at level $2$. We expect that irrational spins are generic in higher levels. 

The freedom in the choice of ground state energy might be considered a defect of the bosonic model. In any case, this freedom is absent from  the 3D ${\cal N}=1$ GS superstring, to which we now turn.  The action may be constructed by first making the replacement  
\be
d\bX^\mu  \to \Pi^\mu  \equiv d\bX^\mu + i \bar\Theta \Gamma^\mu d\Theta
\ee
where $\Gamma^\mu$ are real 3D Dirac matrices and $\Theta$ is a real anticommuting  2-component spinor field, with Majorana conjugate $\bar\Theta = \Theta^T\Gamma^0$. Then we add to the action  a Wess-Zumino term for the supertranslation algebra associated to the closed super-Poincar\'e invariant superspace 3-form  $\Pi^\mu d\bar\Theta\Gamma_\mu d\Theta$. This gives rise to the  following  ``quasi-Hamiltonian'' form of the action
\begin{eqnarray}\label{superstring}
&& S[\bX,\bP;\ell,u] = \int d\tau\oint \!\frac{d\sigma}{2\pi} \left\{\Pi_\tau^\mu {\bP}_\mu     - \frac{1}{2}\ell \left[{\bP}^2 + (T\Pi_\sigma)^2\right]   \right.  \nonumber \\
  &&  \left.  \qquad - u \Pi_\sigma^\mu {\bP}_\mu
+  iT\left( \dot{\bX}^\mu \bar\Theta \Gamma_\mu \Theta^\prime - \bX^\prime{}^\mu \bar\Theta \Gamma_\mu\dot\Theta\right) \right\} \, . 
\end{eqnarray}
where $\Pi_\tau$ and $\Pi_\sigma$ are the $d\tau$ and $d\sigma$ components of the worldsheet 1-form induced by $\Pi$. 
The term linear in $T$ is the WZ term, and we have chosen its coefficient to ensure invariance of the action under the following fermionic gauge invariance (``$\kappa$-symmetry'') with  anticommuting  Majorana spinor parameter $\kappa$:
\begin{eqnarray}
\delta_\kappa\Theta &=& \Gamma_\mu \left(\bP^\mu -T\Pi_\sigma^\mu\right) \kappa\, , \quad
\delta_\kappa \bX^\mu =  -i \bar\Theta \Gamma^\mu\delta_\kappa\Theta\, , \nonumber \\
\delta_\kappa\bP_\mu &=& 2iT \bar\Theta^\prime \Gamma_\mu \delta_\kappa\Theta\, ,  \quad
\delta_\kappa u =  -T \delta_\kappa\ell \, , \nonumber\\
\delta_\kappa\ell &=&   -4i\bar\kappa \left[\dot\Theta + \left(\ell T -u\right)\Theta^\prime\right] \, .
\end{eqnarray}
Observe that  $\Gamma_\mu \left(\bP^\mu -T\Pi_\sigma^\mu\right)$ has zero determinant on the surface defined by the constraints. This means that only one of the two independent components of  $\kappa$ has any effect, so that  only one real component of   $\Theta$ can be `gauged away'.  The Poincar\'e Noether charges are now 
\begin{eqnarray}\label{noether}
{\cal P}_\mu &=& \!\oint\! \frac{d\sigma}{2\pi}  \left\{ {\bP}_\mu + iT \bar\Theta \Gamma_\mu \Theta^\prime\right\} \, ,  \\
{\cal J}^\mu &=&\! \oint\! \frac{d\sigma}{2\pi} \left\{ \left[{\bX}\wedge \left({\bP}+ iT \bar\Theta\Gamma \Theta^\prime\right)\right]
+ \frac{i}{2} \bar\Theta \Theta \left(\bP-T{\bX}^\prime\right)  \right\}^\mu  . \nonumber
\end{eqnarray}
The supersymmetry Noether charges are ($\alpha=1,2$)
\be
{\cal Q}^\alpha = \! \sqrt{2} \oint\! \!\frac{d\sigma}{2\pi} \left\{\left[ \bP_\mu - T{\bX}_\mu^\prime  \right] 
\Gamma^\mu \Theta - \frac{i}{2}\left( \bar\Theta \Theta \right) \Theta^\prime\right\}^\alpha \!\! . 
\ee
The $\kappa$-symmetry variation of all these charges vanishes on the constraint surface. 

To go to the light cone gauge we proceed as before but now we also fix  the $\kappa$-symmetry  by imposing the usual condition \cite{Green:1983wt}
\be\label{fixkap}
\Gamma^+ \Theta=0\, , \qquad \Gamma^\pm = \frac{1}{\sqrt{2}}\left(\Gamma^1\pm \Gamma^0\right)\, . 
\ee
For  the representation $\Gamma^\mu=(i\sigma_2, \sigma_1,\sigma_3)$ of the 3D Dirac matrices, this condition  implies that
\be
\Theta = \left(1/\sqrt{2\sqrt{2}\ p_-}\right)\, \left( \begin{array}{c} \theta \\ 0 \end{array}\right)
\ee
for some anticommuting  worldsheet function $\theta(\tau,\sigma)$. As for the bosonic variables, it is convenient to define
\be
\bar\theta = \theta - \vartheta\, , \qquad \vartheta(\tau) = \frac{1}{2\pi}\oint \! d\sigma  \theta\,  \, . 
\ee
There should be no confusion with the notation for a conjugate spinor as $\theta$ is not a 2-component spinor. Proceeding as before, but now with the additional Fourier expansion 
\be
\bar\theta = \sum_{n=1}^\infty \left[ e^{in\sigma} \theta_n + e^{-in\sigma} \theta_n^*\right]\, ,
\ee
we end up with the  Lagrangian 
\begin{eqnarray}\label{sLag}
L &&\!\!\!\! = \dot x p  + {\dot x}^- p_- + \frac{i}{2}  \vartheta \dot\vartheta 
+\sum_{n=1}^\infty  \left[n^{-1}\left( \dot\alpha_n\alpha_n^* + \dot{\tilde\alpha}_n \tilde\alpha_n^* \right)
+ \theta_n^*\dot\theta_n\right] \nonumber \\
&&\!\!\! \!\!\!\! -\frac{1}{2p_-} \left(p^2 + {\cal M}^2\right) + u_0 \sum_{n=1}^\infty \left( \alpha_n^*\alpha_n - \tilde\alpha_n^* \tilde\alpha_n + n \theta_n^*\theta_n\right)\, , 
\end{eqnarray} 
where
\be
{\cal M}^2 = 2T \sum_{n=1}^\infty \left(\alpha_n^* \alpha_n + \tilde\alpha_n^*\tilde\alpha_n + n\theta_n^*\theta_n\right)\, . 
\ee

The supersymmetry charges in the light-cone gauge are 
\begin{eqnarray}\label{scharges2}
{\cal Q}^1 &=&  \sqrt{\frac{1}{\sqrt{2}\, p_-}}\left\{ p \vartheta + \sqrt{2T} \sum_{n=1}^\infty \left(\alpha_n \theta_n^* + \alpha_n^* \theta_n\right) \right\} \nonumber\\
{\cal Q}^2 &=&  \sqrt{\sqrt{2}\, p_-}\  \vartheta \, . 
\end{eqnarray}
The Lorentz Noether charges in the light-cone gauge are as in (\ref{LorentzString}) but with a different 
expression for $\Lambda$. We will not give this expression here because
the super-Poincar\'e invariant  is not $\Lambda\equiv {\cal P}\cdot{\cal J}$ but rather
\be
\Omega  \equiv {\cal P}\cdot{\cal J} + \tfrac{i}{4} \bar {\cal Q} {\cal Q}\, . 
\ee
To  compute $\Omega$ we need the analog of (\ref{Xminuseq}), which is 
\be
p_- \left(\bar X^-\right)^\prime = - \bar X^\prime P + \tfrac{i}{2}\theta \bar\theta^\prime\, . 
\ee
We again pass over the details; the final result  is 
\be\label{finOm}
\Omega =  \sqrt{2T} \left[\lambda + \tilde\lambda + 
\sum_{n=1}^\infty \frac{i}{n} \left(\alpha_n^* \gamma_n - \alpha_n \gamma_n^*  \right) \right]\, , 
\ee
where $\lambda$ and $\tilde\lambda$ are as before, and 
\be
\!\gamma_n =  \sum_{m=1}^{n-1} (n-m)\theta_m\theta_{n-m}
+ \sum_{m>n}\left(m- \frac{n}{2}\right) \theta_{m-n}^*\theta_m \, . 
\ee
Note that although  the Poincar\'e invariant $\Lambda$ depends on  the fermion zero mode $\vartheta$, this mode cancels from the super-Poincar\'e invariant  $\Omega$.

We quantize as before,  replacing the Grassmann odd  variables by operators obeying canonical anticommutation relations. The non-zero anticommutators are 
\be\label{anticoms}
\left\{\vartheta,\vartheta\right\} \equiv 2\vartheta^2 = 1\, , \qquad  \left\{\theta_n , \theta_n^\dagger\right\} = 1\, . 
\ee
The level matching constraint is now
\be
\tilde N = N + \nu \, , \qquad  \nu = \sum_{n=1}^\infty n\, \theta_n^\dagger \theta_n \, . 
\ee
Taking this into account, one finds that 
\be
\left\{ {\cal Q}^\alpha , {\cal Q}^\beta\right\} = \left(\begin{array}{cc}
\sqrt{2}\, H  & p \\ p & \sqrt{2}\, p_- \end{array} \right)
= \left(\Gamma^\mu \Gamma^0\right)^{\alpha\beta} {\cal P}_\mu\, , 
\ee
provided that the ground state energy is zero, which means that
\be
{\cal M}^2 =4T\left(N+\nu\right) = 4T\tilde N\, .
\ee
The Lorentz charges close  under commutation exactly as for the 3D bosonic string, and it may be verified that they have the expected commutators with the  supersymmetry charges:
$\left[J^\mu , {\cal Q}\right] = -\tfrac{i}{2} \Gamma^\mu {\cal Q}$.

Because of the fermion zero mode, there is a double degeneracy at all levels, and because this zero mode cancels from $\Omega$ the degeneracy is that implied by  the  supersymmetric pairing of one bose with one fermi state, corresponding to the $\pm 1$ eigenspaces of 
$\sqrt{2} \vartheta$. In particular, the ground state is doubly degenerate, and we may interpret the corresponding massless particles in the spectrum as a dilaton and dilatino; note that although 
helicity is not defined for massless states, there is still a distinction between bosons and fermions \cite{Binegar:1981gv,Deser:1991mw}.  This result is consistent with what one would expect from the  low energy effective supergravity of a 3D ${\cal N}=1$ superstring theory because neither the metric nor a 2-form potential propagate modes in this context.  

All higher levels are massive and the states at any given level span an invariant subspace of  $\Omega$, the eigenvalues of which are the superhelicities after division by the mass of the level. The superhelicity is just the average of the two helicities in a supermultiplet, which differ by $\tfrac{1}{2}$, and non-zero superhelicites appear in parity doublets of opposite sign. 
At  level $1$ there are four states, which combine into two massive supermultiplets of superhelicity zero, each of which contains helicities $\pm \tfrac{1}{4}$.  At level $2$ there are 16 states, and hence 8 eigenvalues of $\Omega$. Four are zero, giving four spin-$\tfrac{1}{4}$ particles of helicities $\pm\tfrac{1}{4}$. The four non-zero eigenvalues of $\Omega$ are $(\tfrac{3}{2}, \tfrac{3}{2},-\tfrac{3}{2},- \tfrac{3}{2})$, corresponding to doubly degenerate supermultiplets of helicities  $(\tfrac{7}{4},\tfrac{5}{4})$ and $(-\tfrac{7}{4},-\tfrac{5}{4})$.  At levels $1$ and $2$ we thus find  semion supermultiplets, which were first investigated  in  \cite{Sorokin:1992sy}.  At higher levels we expect generic anyon supermultiplets, see e.g.  \cite{Gorbunov:1997ie}.

There is a classical  3D ${\cal N}=2$ GS superstring and we believe that our quantum results will extend to this case. If so,  there will be four massless  states,  interpretable as a dilaton and axion and their superpartners. Of course,  it remains to be seen whether any of these free quantum 3D strings 
can interact to yield new 3D string theories. 

\noindent
{\bf Acknowledgments}:
We thank Eric Bergshoeff  for helpful discussions, and the Benasque center for Science for  a stimulating environment.  LM wishes to thank DAMTP,  of the University of Cambridge, for hospitality during the early stages of this work. PKT thanks the  EPSRC for financial support. LM acknowlegdes partial support from National Science Foundation Award 0855386.

\end{document}